\documentclass[preprint,showpacs,nofootinbib,prd,aps]{revtex4-1}

\usepackage{graphicx,amstext,amssymb}

\begin{document}

\title{Deciphering  nonfemtoscopic two-pion  correlations in $p+p$
collisions with simple analytical models }

\author{S.V. Akkelin}
\author{ Yu.M. Sinyukov}
\affiliation{Bogolyubov Institute for Theoretical Physics,
Metrolohichna str. 14b, 03680 Kiev,  Ukraine}

\begin{abstract}

A simple model of nonfemtoscopic particle correlations in
proton-proton collisions is proposed. The model takes into account
correlations induced by the conservation laws as well as
correlations induced by minijets. It reproduces well the two-pion
nonfemtoscopic correlations of like-sign and unlike-sign pions in
proton-proton collision events at $\sqrt{s} = 900$ GeV analyzed by
the ALICE Collaboration. We also argue that similar nonfemtoscopic
correlations can appear in the hydrodynamic picture with
event-by-event fluctuating nonsymmetric initial conditions that are
typically associated with nonzero higher-order flow harmonics.
\end{abstract}

\pacs{13.85.Hd, 25.75.Gz}
 \maketitle

\section{Introduction}

The method of identical particle correlation femtoscopy   gives us
the  possibility to measure  length and time scales with an accuracy
of more than  $10^{-14}$ m and $10^{-22}$   sec, respectively.   It
is widely used now for  the determination of sizes and lifetimes of
the sources of  particle emission such as  the systems created in
heavy ion, hadron, and lepton collisions (for reviews see, e.g.,
Ref. \cite{Pratt}). The method is grounded on the Bose-Einstein or
Fermi-Dirac symmetry properties of  quantum states and, in fact,
measures correlations between numbers of identical particles with
close energy and momentum.

In the method of correlation femtoscopy the space-time structure of
the systems  is usually represented in terms of the interferometry
radii that are the result of a Gaussian fit of a two-particle
correlation function depending on the momentum difference in the
pair. For bosons such correlations, reflecting spatiotemporal scales
of an extended source, are caused by the quantum Bose-Einstein
statistics. These correlation functions can be  obtained from  the
ratio of the two- (identical) particle momentum spectra to the
product of the single particle ones. The former typically need to be
corrected for the Coulomb and strong final state interactions (FSI)
that depend on  space-time points of the last collisions for the
detected pair (see, e.g., the recent review \cite{Ledn} and
references therein). In pioneering  papers \cite{Kopylov} the
measured interferometry radii were interpreted as
 geometrical sizes of the systems. Later on it was found
\cite{Gyulassy,sin1} that because the typical systems formed in
experiments with heavy ion collisions are expanding and have
inhomogeneous structures, the above geometrical interpretation is
not complete, and the interpretation of  interferometry radii as
homogeneity lengths in the systems was proposed \cite{sin2}. It was
demonstrated that the homogeneity lengths carry information about
the dynamical properties (rate of expansion, lifetime, etc.) of the
source and depend on the mean momentum of pairs \cite{sin1,sin2}. In
particular, it is firmly established that specific transverse
momentum dependence of  femtoscopy scales - interferometry radii -
in heavy ion collisions is mostly caused by a collective
(hydrodynamical) transverse expansion of the systems formed in these
collisions \cite{Pratt}.

The situation is more complicated, and the above method has to be
modified  for elementary particle collisions, like $p+p$, which have
smaller spatiotemporal scales as compared to heavy ion collisions.
It became clear \cite{Alice,Alice1,Star} that for relatively small
systems the additional two-particle correlations affect the
correlation functions in the kinematic region where quantum
statistical (QS)  and FSI correlations are usually observed. The
well-known  example of such additional correlations is the
correlation induced by  total energy and momentum conservation laws
(see, e.g., Ref. \cite{Lisa}). As opposed to the QS and FSI
correlations, which are familiar  from the correlation femtoscopy
method, and so are sometimes called   femtoscopy correlations, these
correlations are not directly related to the spatiotemporal scales
of the emitter and are therefore called nonfemtoscopic correlations.
Since the latter noticeably affect  correlation functions for small
systems, the interferometry radii extracted from the complete
correlation function in $p+p$ collisions depend strongly on the
assumption about the so-called correlation baseline - the strength
and momentum dependence of the nonfemtoscopic correlations
\cite{Alice,Alice1,Star}.   It has an influence on the
interpretation of the momentum dependence of the interferometry
radii in $p+p$ collisions, where the possibility of hydrodynamic
behavior of matter is questionable. Therefore, for successful and
unambiguous applications of the correlation femtoscopy method to
elementary particle collisions, one needs to know the mechanisms  of
nonfemtoscopic correlations to separate the femtoscopic  and
nonfemtoscopic correlations.

Recently, the ALICE Collaboration utilized some event generators,
which  do not include  effects of quantum statistics, for an
estimate of the correlation baseline (i.e., nonfemtoscopic
correlation function of identical pions) under the Bose-Einstein
peak at  LHC energies \cite{Alice,Alice1}. It was motivated by a
reasonable agreement of the corresponding event generator
simulations with the experimental data for  correlation functions of
oppositely charged pions in $p+p$ collisions at the same energy
\cite{Alice,Alice1}. The calculated correlation baseline  has been
utilized by the ALICE Collaboration to extract  femtoscopic
correlations from measured identical pion two-particle correlation
functions \cite{Alice}.

Because  the utilized event generators  account for energy-momentum
conservation and   emission of minijets, it was conjectured in Refs.
\cite{Alice,Alice1} that some specific  peculiarities of the
unlike-sign pion correlations as well as like-sign nonfemtoscopic
pion correlations can be caused by the jetlike and energy-momentum
conservation induced correlations. In what follows, we support this
conjecture. We develop  a simple analytical model with the minimal
number of parameters for the two-pion  correlations induced by
minijets and transverse momentum conservation law, and show that
this model can fit the correlations of unlike-sign pion pairs at
$\sqrt{s} = 900$ GeV $p+p$ collisions measured by  the ALICE
Collaboration \cite{Alice}. Also, with a reasonable change of
parameters,  the model can  fit  nonfemtoscopic correlations of
like-sign pion pairs obtained from the event generator simulations
of $p+p$ collision events at the same energy that are reported in
Ref. \cite{Alice}.\footnote{For convenience, we compare results of
our model with the PHOJET event generator \cite{PHOJET} simulations
reported in Ref. \cite{Alice}. Note that the simulations carried out
by the ALICE Collaboration gave similar results for all utilized
event generators \cite{Alice,Alice1}.} Our model is simple and
analytical, and clearly demonstrates the interplay between minijet
 and conservation law induced correlations in the formation of the
nonfemtoscopic correlations.

But this is not  the whole  story. Despite the fact that the event
generators do reproduce the unlike pion correlations as has been
verified by the ALICE Collaboration, there is enough room to doubt
whether the nonfemtoscopic correlations of like-sign pions are
properly simulated by them. Indeed, none of the utilized event
generators  can reproduce the LHC data on the multiplicities and
momentum spectra well (see, e.g., \cite{Alice-rev} and references
therein). This suggests that some essential ingredients may be
missing in these event generators. The possible candidate for missed
dynamics is hydrodynamics. The latter describes well the dynamics of
heavy ion collisions, see, e.g.,  Ref. \cite{Heinz-1} and references
therein. If hydrodynamics is applied also for $p+p$ collisions with
high multiplicities \cite{Landau}, then the other mechanisms of
nonfemtoscopic correlations should  be taken into account. In what
follows, we present some heuristic arguments, based on an
illustrative analytical model, that unlike-sign two-pion correlation
functions calculated in hydrodynamics with event-by-event
asymmetrically fluctuating initial densities can be qualitatively
similar at relatively low $q_{inv}$ to the ones calculated in
PHOJET-like event generators, where these correlations for
relatively low $q_{inv}$ are mainly caused by the minijets. Finally,
we briefly discuss what these results can mean for modeling of the
correlation baseline (i.e., identical pion nonfemtoscopic
correlations) and for applications of the correlation femtoscopy
method to $p+p$ collisions.

\section{Definitions and parametrizations of two-particle correlations}

The two-particle  correlation function is defined as
\begin{eqnarray}
C(p_{1},p_{2})=\frac{P(p_{1},p_{2})}{P(p_{1})P(p_{2})}, \label{1}
\end{eqnarray}
where $P(p_{1},p_{2})$ is the probability of observing two particles
with momenta $\textbf{p}_{1}$ and $\textbf{p}_{2}$, while $P(p_{1})$
and $P(p_{2})$ designate the single-particle probabilities.
Experimentally, the two-particle correlation function  is defined as
the ratio of the distribution of particle  pairs from the same
collision event to the   distribution of pairs with particles taken
from different events.  In heavy ion collisions almost all the
correlations between identical pions with low relative momentum are
due to quantum statistics  and final state interactions. In this
case
\begin{eqnarray}
C(p_{1},p_{2})=C_{F}(\textbf{p},\textbf{q}), \label{3}
\end{eqnarray}
where  ${\bf p}=({\bf p}_{1}+{\bf p}_{2})/2$, ${\bf q}={\bf
p}_{2}-{\bf p}_{1}$, and $C_{F}$ denotes  the  femtoscopic
correlation function. For identical bosons $C_{F}$ is often
parameterized (after corrections for FSI correlations)  by the
Gaussian form which for the one-dimensional parametrization looks
like
\begin{eqnarray}
 C_{F}(|{\bf
p}|,q_{inv})=1+\lambda \exp{(-R^{2}_{inv}q^{2}_{inv})}. \label{4}
\end{eqnarray}
Here  $\lambda$    describes the correlation strength, $R_{inv}$ is
the Gaussian "invariant" interferometry  radius, and
$q_{inv}=\sqrt{({\bf{p}}_{2}-{\bf{p}}_{1})^{2}-(E_{2}-E_{1})^{2}}$
is equal to  the modulus of the 3-momentum difference in the pair
rest frame.

In elementary particle collisions additional (nonfemtoscopic)
correlations,   like those arising from jet/string fragmentation and
from energy and momentum conservation (see, e.g., Refs.
\cite{Alice,Alice1,Star}) can also give a significant contribution.
Then, assuming the factorization property,
\begin{eqnarray}
C(p_{1},p_{2})=C_{F}(\textbf{p},\textbf{q})C_{NF}(\textbf{p},\textbf{q}).
\label{4-1}
\end{eqnarray}
Here $C_{NF}$ denotes the  nonfemtoscopic correlation function, and
in the simplest case the nonfemtoscopic effects can be parametrized
as, e.g., 2nd order polynomial,
\begin{eqnarray}
C_{NF}(|{\bf p}|,q_{inv})=a+bq_{inv}+cq_{inv}^{2}.
 \label{6}
\end{eqnarray}
This form can be used together with some parametrization of $C_{F}$
(e.g., with (\ref{4})) in order to fit the correlation function
$C(p_{1},p_{2})$ for small systems, as has been done, for example,
by the STAR Collaboration  for two-pion correlation functions in
$p+p$ collisions at $\sqrt{s}=200$ GeV \cite{Star}. At $c>0$ the
phenomenological parametrization (\ref{6}) explicitly reproduces the
well known effect of positive correlations between particles with
large relative momenta  $|\textbf{q}|$ caused by the energy-momentum
conservation laws; see the energy and momentum conservation-induced
correlation model for $C_{NF}$ \cite{Lisa}. Note that $a$, $b$, and
$c$ in Eq. (\ref{6}) depend, in general, on $|{\bf p}|$.

\section{ Probability densities  of
 distinguishable equivalent particles}

Aiming to estimate the nonfemtoscopic pion correlations, we consider
emitted  pions as distinguishable, yet  equivalent noninteracting
particles with symmetrical probability density functions, thereby,
excluding femtoscopic QS and FSI correlations. To a certain  extent
this corresponds to the (quasi)classical approximation used in
current event generators, like  PHOJET.

Let us assume that $N$ bosons  of the same species (say, pions) are
produced with momenta $\textbf{p}_{1},...,\textbf{p}_{N}$ in $(N+X)$
multiparticle production events. Then the $N$-particle probability
density $P_{N}(p_{1},...,p_{N})$ is a symmetrical function  for all
$N!$ permutations of the particle momenta $p_{i}$. For convenience,
it is normalized by
\begin{eqnarray}
\int d\Omega_{p} P_{N}(p_{1},...,p_{N})=1,
 \label{8n}
\end{eqnarray}
where $d\Omega_{p} =  \frac{d^{3}p_{1}}{E_{1}} ...
\frac{d^{3}p_{N}}{E_{N}}$. Then the $N'$-pion probability density,
$N'<N$, is defined as
\begin{eqnarray}
P_{N}(p_{1},...,p_{N'})= \int d\Omega_{p^{*}}
E_{i}^{*}\delta^{(3)}(\textbf{p}_{1}-\textbf{p}_{1}^{*})...E_{N'}^{*}\delta^{(3)}(\textbf{p}_{N'}-\textbf{p}_{N'}^{*})P_{N}(p_{1}^{*},...,p_{N}^{*}).
 \label{def-n}
\end{eqnarray}
In what follows, we use an assumption of distinguishability  of
equivalent particles which  means the  absence  of quantum
interference between possibilities corresponding  to all $N!$
permutations of the particle momenta $p_{i}$. Therefore the
symmetrized $N$-particle probability density is defined as
\begin{eqnarray}
P_{N}(p_{1},...,p_{N})=\frac{1}{N!} \sum_{i \neq  ... \neq k=1}^{N}
\int d\Omega_{p^{*}}
E_{i}^{*}\delta^{(3)}(\textbf{p}_{1}-\textbf{p}_{i}^{*})...E_{k}^{*}\delta^{(3)}(\textbf{p}_{N}-\textbf{p}_{k}^{*})\widehat{P}_{N}(p_{1}^{*},...,p_{N}^{*}),
 \label{10n}
\end{eqnarray}
where the nonsymmetrized $N$-particle probability density
$\widehat{P}_{N}(p_{1},...,p_{N})$ is normalized to unity, and
 $N!$ in the denominator is  required  to
guarantee the normalization  condition (\ref{8n}). Taking into
account Eq. (\ref{def-n}), we see that the single-particle
probability $P_{N}(p_{1})$ and the two-particle probability
$P_{N}(p_{1},p_{2})$  can be written as
\begin{eqnarray}
P_{N}(p_{1})= \frac{1}{N} \sum_{i=1}^{N} \int d\Omega_{p^{*}}
E_{i}^{*}\delta^{(3)}(\textbf{p}_{1}-\textbf{p}_{i}^{*})\widehat{P}_{N}(p_{1}^{*},...,p_{N}^{*}),
 \label{7n-1} \\
P_{N}(p_{1},p_{2})= \frac{1}{N(N-1)}\sum_{i \neq j=1}^{N}\int
d\Omega_{p^{*}}
E_{i}^{*}\delta^{(3)}(\textbf{p}_{1}-\textbf{p}_{i}^{*})E_{j}^{*}\delta^{(3)}(\textbf{p}_{2}-\textbf{p}_{j}^{*})\widehat{P}_{N}(p_{1}^{*},...,p_{N}^{*}).
 \label{7n-2}
\end{eqnarray}
The nonsymmetrized $N$-pion probability density  in such events
reads
\begin{eqnarray}
\widehat{P}_{N}(p_{1},...,p_{N})= \frac{1}{K}\sum_{X} \int
d\Omega_{k}\delta^{(4)} (p_{a}+p_{b}-\sum_{i =1}^{N}p_{i}-\sum_{j
=1}^{X}k_{j}) |M_{N+X}(p_{1},...,k_{X})|^{2},
 \label{11nn}
\end{eqnarray}
where $M_{N+X}(p_{1},...,k_{X})$ is the  nonsymmetrized
$(N+X)$-particle production amplitude, $p_{a}$ and $p_{b}$ are
4-momenta of colliding particles (protons in $p+p$ collision
events),  and $K$ is the normalization factor,
\begin{eqnarray}
K=\sum_{X} \int d\Omega_{k}d\Omega_{p}\delta^{(4)}
(p_{a}+p_{b}-\sum_{i =1}^{N}p_{i}-\sum_{j =1}^{X}k_{j})
|M_{N+X}(p_{1},...,k_{X})|^{2}.
 \label{12nn}
\end{eqnarray}

Expression  (\ref{11nn}) for $\widehat{P}_{N}(p_{1},...,p_{N})$ is
rather complicated  because, in particular,  it depends on $X$
particles that are produced in addition to $N$ pions. Production of
 additional particles also means  that one can hardly expect
that the total energy or momentum of the pion subsystem are
constants in the system's center of mass; instead, one can expect
that they fluctuate from event to event. Then, motivated by Eq.
(\ref{11nn}), we assume that a nonsymmetrized $N$-pion probability
density can be written as
\begin{eqnarray}
\widehat{P}_{N}(p_{1},...,p_{N})= \frac{1}{K} \delta
(p_{1},...,p_{N}) F_{N}(p_{1},...,p_{N}),
 \label{11n}
\end{eqnarray}
where $F_{N}(p_{1},...,p_{N})$ is a nonsymmetrized function of
pionic momenta and $\delta(p_{1},...,p_{N})$ denotes the average
constraints on the $N$-pion states that appear due to energy and
momentum conservation in multiparticle production events. Then the
normalization factor is
\begin{eqnarray}
K=\int d\Omega_{p} \delta(p_{1},...,p_{N}) F_{N}(p_{1},...,p_{N}).
 \label{12n}
\end{eqnarray}
If  the only correlations are the correlations  associated  with
energy and momentum conservation, we have
\begin{eqnarray}
F_{N}(p_{1},...,p_{N}) = f(p_{1})f(p_{2})...f(p_{N-1})f(p_{N}),
 \label{14n}
\end{eqnarray}
and calculations in the large $N$ limit of single-particle and
two-particle probability densities  result in the special case of
the EMCIC parametrization \cite{Lisa} of the correlations induced by
the energy and momentum conservation laws. However,  such a simple
prescription cannot result in the unlike-sign pion correlations
measured by the ALICE Collaboration \cite{Alice}. Also it does not
reproduce the nonfemtoscopic like-sign pion correlations generated
by the PHOJET event generator \cite{Alice}.

\section{Analytical model for the two-pion  correlations induced by minijets
and  momentum conservation}

The ALICE Collaboration has analyzed the unlike-sign pion
correlations \cite{Alice} and found that they can be well reproduced
by event generators which account for, among others factors, total
energy-momentum conservation  and minijet production. In what
follows, we assume that just these two factors induce the observed
behavior of the unlike-sign pion correlations  and are responsible
for the nonfemtoscopic correlations  in like-sign ones. In our
simple model we start from zero total transverse momentum of the
system, keeping in mind that for a subsystem this statement should
be weakened. Also, we neglect the constraints conditioned by the
conservation of energy and longitudinal momentum, supposing that the
system under consideration is an $N$-pion subsystem in a small
midrapidity region of the total system. Then in the first
approximation
 \begin{eqnarray}
\delta(p_{1},...,p_{N})=\delta^{(2)}({\bf p}_{T1} + {\bf p}_{T2} +
... +{\bf p}_{TN}),
 \label{13n}
\end{eqnarray}
where ${\bf p}_{T1},{\bf p}_{T2}, ... {\bf p}_{TN}$ are transverse
components of the momenta of the $N$ particles.

Let us assume that there are no other correlations in the production
of $N$-pion states except the correlations induced by the transverse
momentum conservation and cluster (minijet) structures in momentum
space.  For the sake of simplicity we assume here that only the
two-particle clusters appear.  Then one can write for fairly large
$N \gg 1$
\begin{eqnarray}
F_{N}(p_{1},...,p_{N}) =
 f(p_{1})...f(p_{N})Q(p_{1},p_{2})...Q(p_{N-1},p_{N}),
  \label{15n}
\end{eqnarray}
where $Q(p_{i},p_{j})$ denotes the jetlike correlations between
momenta  $\textbf{p}_{i}$ and $\textbf{p}_{j}$; the existence of
such correlations means that  $F_{N}$ cannot be expressed as a
product of one-particle distributions. Then, utilizing the integral
representation of the $\delta$-function by means of the Fourier
transformation, $\delta^{(2)}({\bf p}_{T})=(2 \pi)^{-2}\int
d^{2}r_{T}\exp(i{\bf r}_{T}{\bf p}_{T})$, and accounting for Eqs.
(\ref{7n-1}), (\ref{11n}), (\ref{13n}), (\ref{15n}), the
single-particle probability reads
\begin{eqnarray}
P_{N}(p_{1})= \frac{1}{(2 \pi)^{2}K} \int d^{2}r_{T}  G_{N}({\bf
p}_{1}, {\bf r}_{T}),
 \label{16n}
\end{eqnarray}
where
\begin{eqnarray}
 G_{N}({\bf p}_{1}, {\bf r}_{T}) =  \int d\Omega_{p^{*}}
E_{1}^{*}\delta^{(3)}(\textbf{p}_{1}-\textbf{p}_{1}^{*}) e^{i {\bf
r}_{T} ( {\bf p}_{T1}^{*} + ...+ {\bf p}_{TN}^{*})}
F_{N}(p_{1}^{*},...,p_{N}^{*}).
 \label{17n}
\end{eqnarray}
A possibility of different  cluster configurations of particles
means, in particular, that registered particles with momenta
$\textbf{p}_1$ and $\textbf{p}_2$ can belong either to different
minijets or to the same minijet. Then, taking into account Eqs.
(\ref{7n-2}), (\ref{11n}), (\ref{13n}), (\ref{15n}), we get
\begin{eqnarray}
P_{N}(p_{1}, p_{2})= \frac{N}{N(N-1)}  P_{N}^{1jet}(p_{1}, p_{2}) +
\frac{N(N-1)-N}{N(N-1)} P_{N}^{2jet}(p_{1}, p_{2}),
   \label{18n}
\end{eqnarray}
where
\begin{eqnarray}
P_{N}^{1jet}(p_{1}, p_{2})= \frac{1}{(2 \pi)^{2}K} \int d^{2}r_{T}
G_{N}^{1jet}({\bf p}_{1},
{\bf p}_{2},{\bf r}_{T}),  \label{9.1} \\
P_{N}^{2jet}(p_{1}, p_{2})= \frac{1}{(2 \pi)^{2}K} \int d^{2}r_{T}
G_{N}^{2jet}({\bf p}_{1}, {\bf p}_{2},{\bf r}_{T}),  \label{19n}
\end{eqnarray}
and
\begin{eqnarray}
G_{N}^{1jet}({\bf p}_{1}, {\bf p}_{2},{\bf
 r}_{T}) = \int d\Omega_{p^{*}}
E_{i}^{*}\delta^{(3)}(\textbf{p}_{1}-\textbf{p}_{1}^{*})E_{j}^{*}\delta^{(3)}(\textbf{p}_{2}-\textbf{p}_{2}^{*})
e^{i {\bf r}_{T} ( {\bf p}_{T1}^{*} + ...+ {\bf p}_{TN}^{*})} F_{N},
 \label{20n} \\
 G_{N}^{2jet}({\bf p}_{1}, {\bf p}_{2},{\bf r}_{T}) =  \int d\Omega_{p^{*}}
E_{i}^{*}\delta^{(3)}(\textbf{p}_{1}-\textbf{p}_{1}^{*})E_{j}^{*}\delta^{(3)}(\textbf{p}_{2}-\textbf{p}_{3}^{*})
e^{i {\bf r}_{T} ( {\bf p}_{T1}^{*} + ...+ {\bf p}_{TN}^{*})} F_{N}.
 \label{21n}
\end{eqnarray}
Here $F_{N} \equiv F_{N}(p_{1}^{*},...,p_{N}^{*})$.
 The first term on the right-hand side of Eq. (\ref{18n}) is
associated with events where the two registered particles belong to
the same minijet, and the second term corresponds to events where
the particles are  from different minijets. Evidently, the former is
relatively rare; however, notice that the  first term can be
significant for small systems with not very large $N$.

Now let us check whether this model can reproduce, with reasonable
parameters, the correlation functions of unlike-sign pions measured
by the  ALICE Collaboration \cite{Alice} and nonfemtoscopic
correlations of like-sign pions that are generated in the PHOJET
simulations and utilized as the correlation baseline by the ALICE
Collaboration \cite{Alice}. Calculations within the  model will
deliberately be as simple as possible just to demonstrate its
viability.  Here we do not use  the approximate methods like the
saddle point approach; instead, we utilize appropriate analytical
parametrizations  of the functions of interest, namely,
\begin{eqnarray}
f(p_{i})=E_{i}\exp{\left (-\frac{{\bf
p}_{i,T}^{2}}{T_{T}^{2}}\right)} \exp{\left (-\frac{{\bf
p}_{i,L}^{2}}{T_{L}^{2}}\right )},
  \label{14}
\end{eqnarray}
and
\begin{eqnarray}
Q(p_{i},p_{j}) = \exp{\left(-\frac{({\bf p}_{i}-{\bf
p}_{j})^{2}}{\alpha^2} \right )},
  \label{14.0}
\end{eqnarray}
where $T_{T}$, $T_{L}$, and $\alpha$  are some parameters, and in
what follows, we assume that $T_{L} \gg T_{T}$.  In accordance with
the ALICE baseline obtained from the PHOJET event generator
simulations, we assume that only $q_{inv}$ is measured for each
$\textbf{p}_T$ bin.  Assuming  that longitudinal components of the
registered particles are equal to zero, $p_{1L}=p_{2L}=0$, we
approximate $q_{inv}^{2}$ as
\begin{eqnarray}
q_{inv}^{2} \approx {\bf q}_{T}^{2} \left ( \frac {m^{2}+{\bf
p}_{T}^{2} \sin^{2}{\phi}} {m^{2}+{\bf p}_{T}^{2}} \right ),
  \label{15}
\end{eqnarray}
where  $\phi$ denotes the unregistered  angle between ${\bf p}_{T}$
and ${\bf q}_{T}$, $ {\bf p}_{T} {\bf q}_{T} = |{\bf p}_{T}| |{\bf
q}_{T}| \cos {\phi}$.  Then
\begin{eqnarray}
C_{NF}(|{\bf p}_{T}|,q_{inv})=\frac {\int_{0}^{2 \pi}d\phi
P_{N}(p_{1}, p_{2}) }{\int_{0}^{2 \pi}d\phi
P_{N}(p_{1})P_{N}(p_{2})},
  \label{15.1}
\end{eqnarray}
and, taking into account Eq. (\ref{18n}), we get
\begin{eqnarray}
C_{NF}(|{\bf p}_{T}|,q_{inv})=\frac{N-2}{N-1} \left (
C_{N}^{2jet}(|{\bf p}_{T}|,q_{inv}) +
\frac{1}{N-2}C_{N}^{1jet}(|{\bf p}_{T}|,q_{inv})\right ),
  \label{16}
\end{eqnarray}
where
\begin{eqnarray}
C_{N}^{2jet}(|{\bf p}_{T}|,q_{inv})= \frac {\int_{0}^{2 \pi}d\phi
P_{N}^{2jet}(p_{1}, p_{2}) }{\int_{0}^{2
\pi}d\phi P_{N}(p_{1})P_{N}(p_{2})},  \label{17-00} \\
C_{N}^{1jet}(|{\bf p}_{T}|,q_{inv})= \frac {\int_{0}^{2 \pi}d\phi
P_{N}^{1jet}(p_{1}, p_{2}) }{\int_{0}^{2 \pi}d\phi
P_{N}(p_{1})P_{N}(p_{2})}.
  \label{17}
\end{eqnarray}

 It is well known (see, e.g., Ref. \cite{Lisa}) that the influence
of exact conservation laws on single-particle and two-particle
momentum probability densities at the  $N$-particle production
process depends on the value of $N$ and disappears  at $N\rightarrow
\infty$. Since one considers a subsystem of $N$ pions but not the
total system, to weaken the influence of the total transverse
momentum conservation on pions we shall consider $C_{M}^{1jet}$ and
$C_{M}^{2jet}$ with $M>N$  instead of $C_{N}^{1jet}$ and
$C_{N}^{2jet}$ in Eq. (\ref{16}). This is the simplest way to
account for a weakened conservation law in our model. At the same
time, the factor $1/(N-2)$ in (\ref{16}) remains the same since it
is associated with the combinatorics of the distribution of
particles between clusters in momentum space ("minijets"), which
happens whether or not one weakens the total momentum conservation
law. Also, for more exact fitting of the data points in each average
transverse momentum bin, we utilize the auxiliary factors $\Lambda$.
 Then Eq. (\ref{16}) gets the form
\begin{eqnarray}
C_{NF}(|{\bf p}_{T}|,q_{inv})=\Lambda (|{\bf p}_{T}|) (
C_{M}^{2jet}(|{\bf p}_{T}|,q_{inv}) +
\frac{1}{N-2}C_{M}^{1jet}(|{\bf p}_{T}|,q_{inv}) ).  \label{data-NF}
\end{eqnarray}

\begin{figure}[h]
\centering
\includegraphics[scale=1.0]{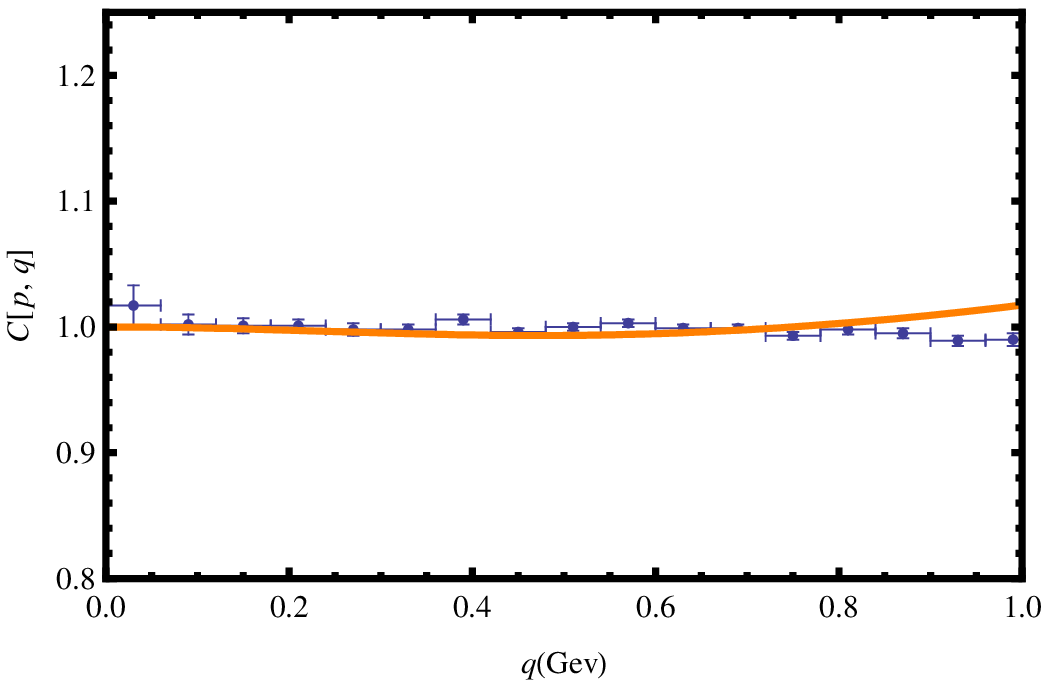}
\caption{\label{fig1} The nonfemtoscopic  correlation functions of
like-sign pions in the $0.1<p_{T}<0.25$ GeV bin from a simulation
using PHOJET \cite{Alice,data} (solid dots) and those calculated
from the analytical model: minijets $+$ momentum conservation (solid
line). See the text for details.}
\end{figure}

\begin{figure}[h] \centering
\includegraphics[scale=1.0]{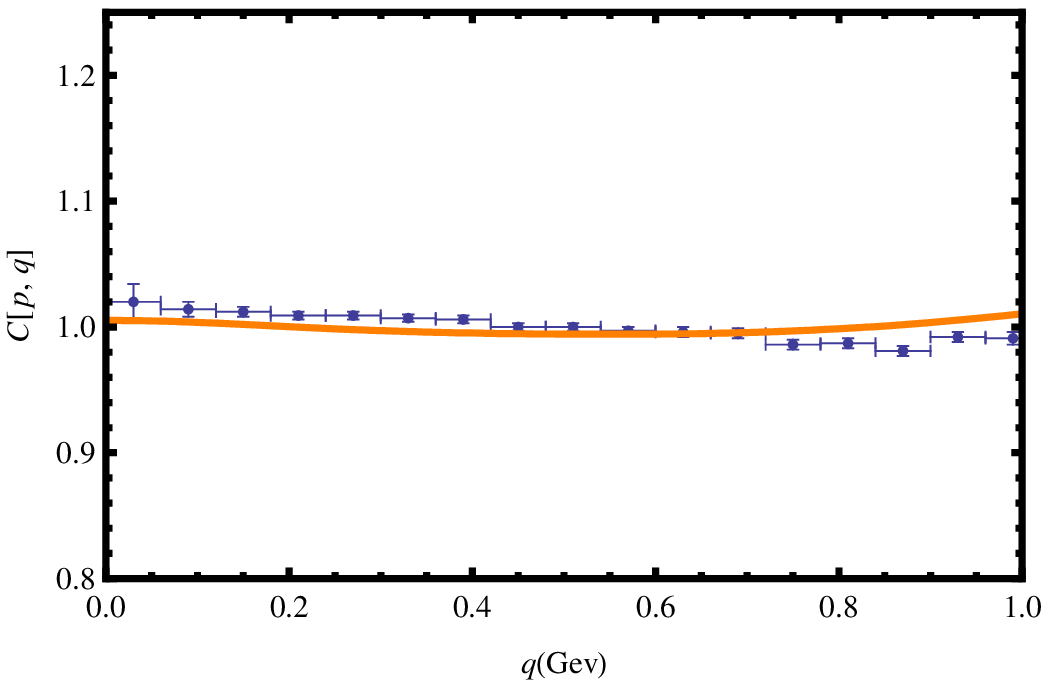}
\caption{\label{fig2}The same as Fig. \ref{fig1} but in the
$0.25<p_{T}<0.4$ GeV bin.}
\end{figure}

\begin{figure}[h]
\centering
\includegraphics[scale=1.0]{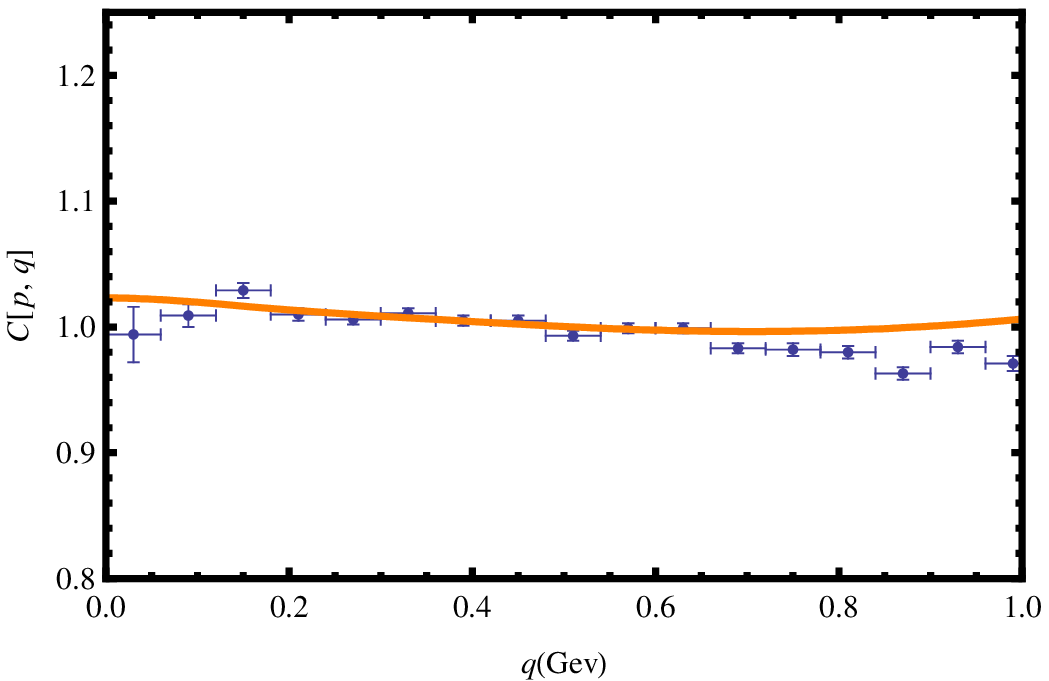}
\caption{\label{fig3}The same as Fig. \ref{fig1} but in the
$0.4<p_{T}<0.55$ GeV bin.}
\end{figure}

\begin{figure}[h]
\centering
\includegraphics[scale=1.0]{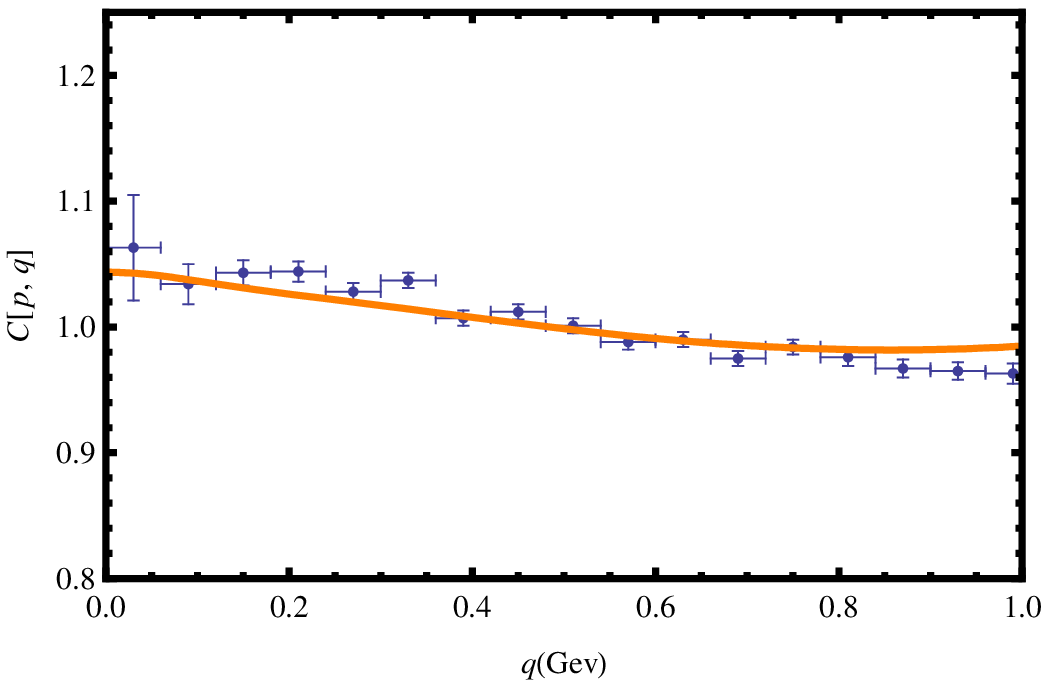}
\caption{\label{fig4}The same as Fig. \ref{fig1} but in the
$0.55<p_{T}<0.7$ GeV bin.}
\end{figure}

\begin{figure}[h]
\centering
\includegraphics[scale=1.0]{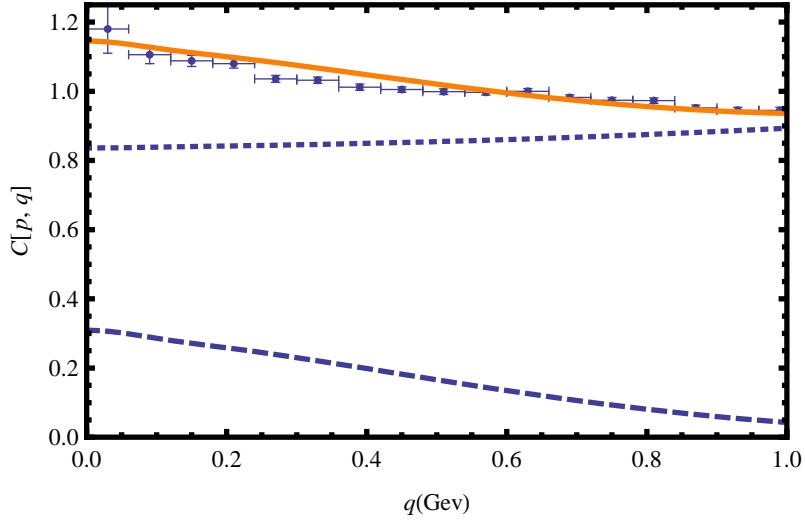}
\caption{\label{fig5} The nonfemtoscopic correlation functions of
like-sign pions in the $0.7<p_{T}<1.0$ GeV bin from a simulation
using PHOJET \cite{Alice,data} (solid dots) and those calculated
from the analytical model (solid line).  The  contributions to the
nonfemtoscopic correlation function from the first term of Eq.
(\ref{data-NF})
 (dotted line) and from the second one (dashed line) are also presented.}
\end{figure}

\begin{figure}[h]
\centering
\includegraphics[scale=1.0]{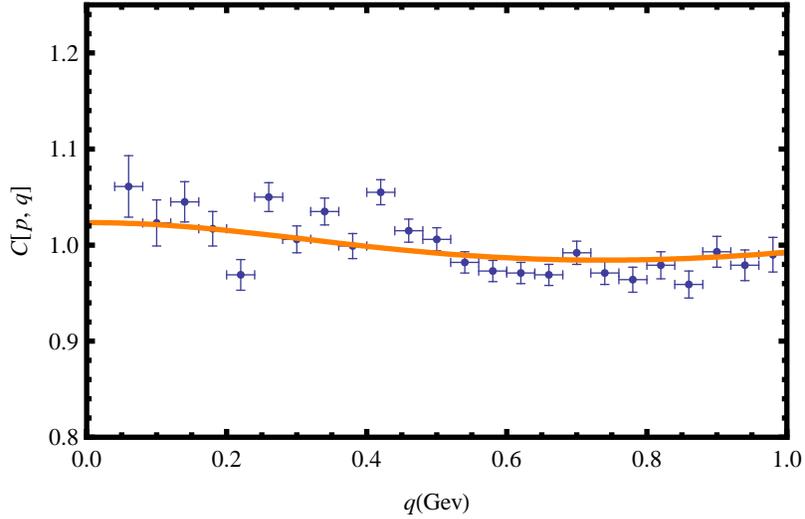}
\caption{\label{ffig1} The  correlation functions of unlike-sign
pions in the $0.1<p_{T}<0.25$ GeV bin measured by the ALICE
Collaboration  from Refs. \cite{Alice,data} (solid dots) and those
calculated from  the analytical model: minijets $+$ momentum
conservation (solid line). See the text for details.}
\end{figure}

\begin{figure}[h] \centering
\includegraphics[scale=1.0]{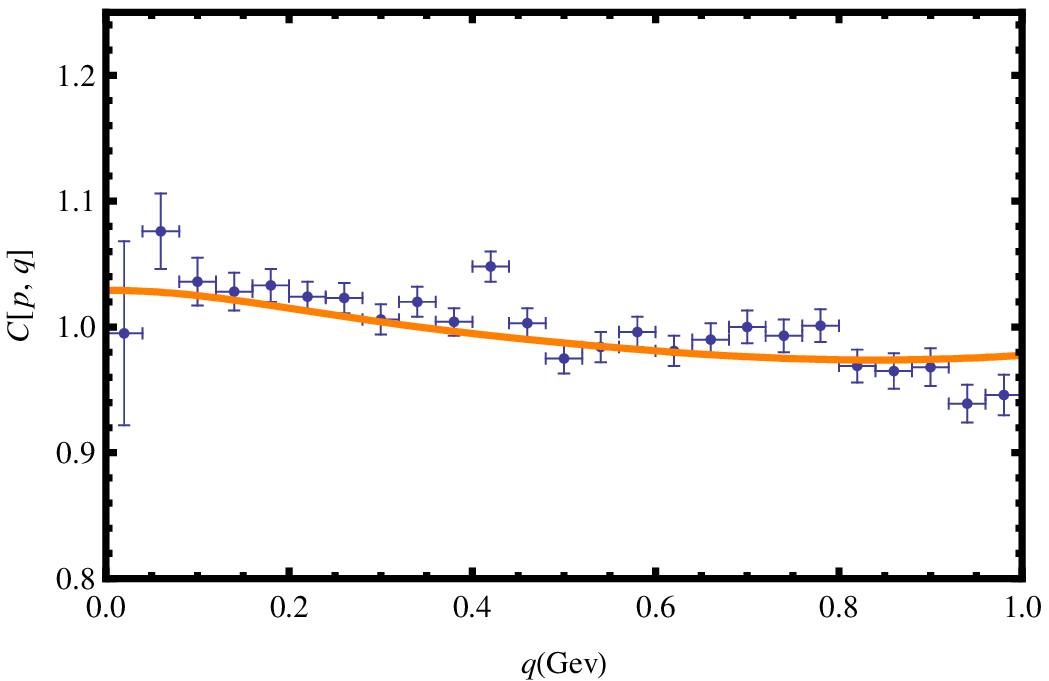}
\caption{\label{ffig2}The same as Fig. \ref{ffig1} but in the
$0.25<p_{T}<0.4$ GeV bin.}
\end{figure}

\begin{figure}[h]
\centering
\includegraphics[scale=1.0]{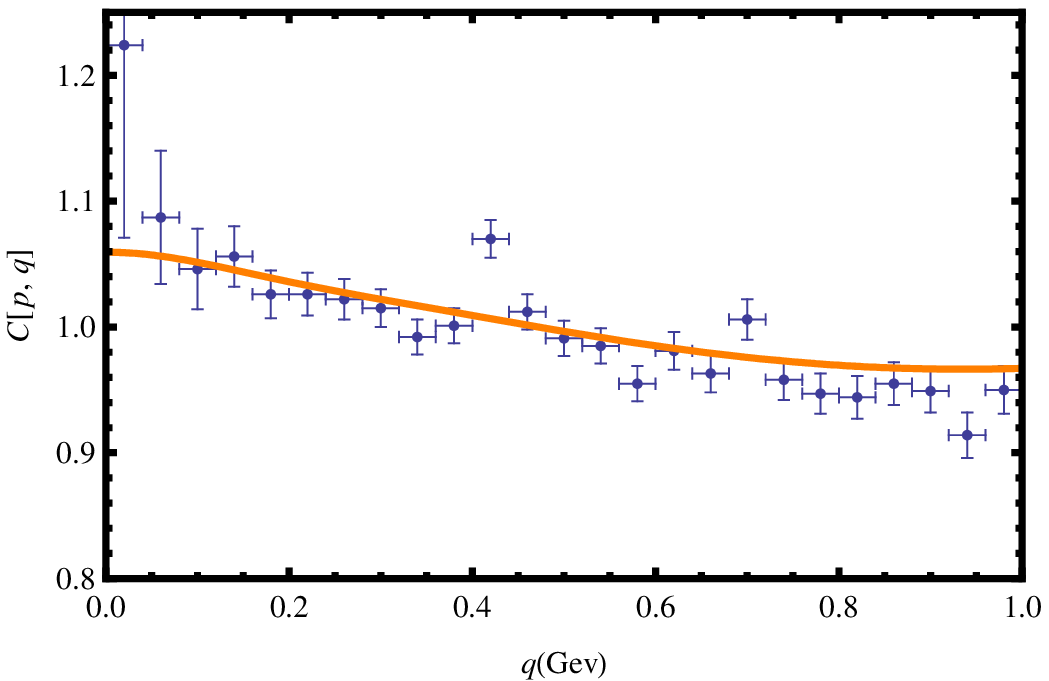}
\caption{\label{ffig3}The same as Fig. \ref{ffig1} but in the
$0.4<p_{T}<0.55$ GeV bin.}
\end{figure}

\begin{figure}[h]
\centering
\includegraphics[scale=1.0]{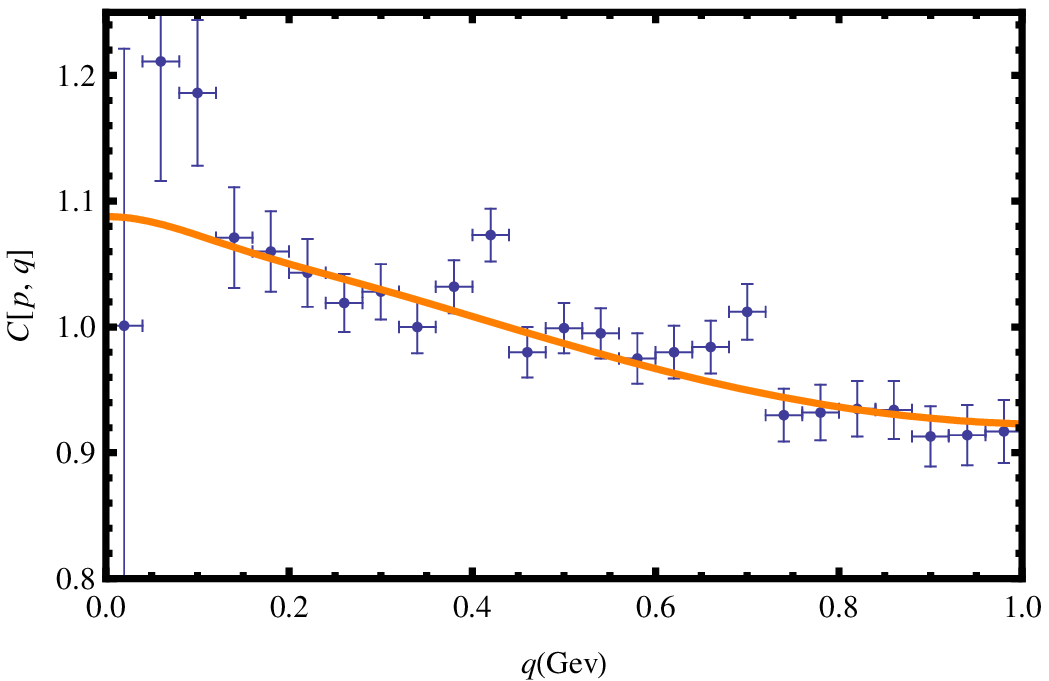}
\caption{\label{ffig4}The same as Fig. \ref{ffig1} but in the
$0.55<p_{T}<0.7$ GeV bin.}
\end{figure}

\begin{figure}[h]
\centering
\includegraphics[scale=1.0]{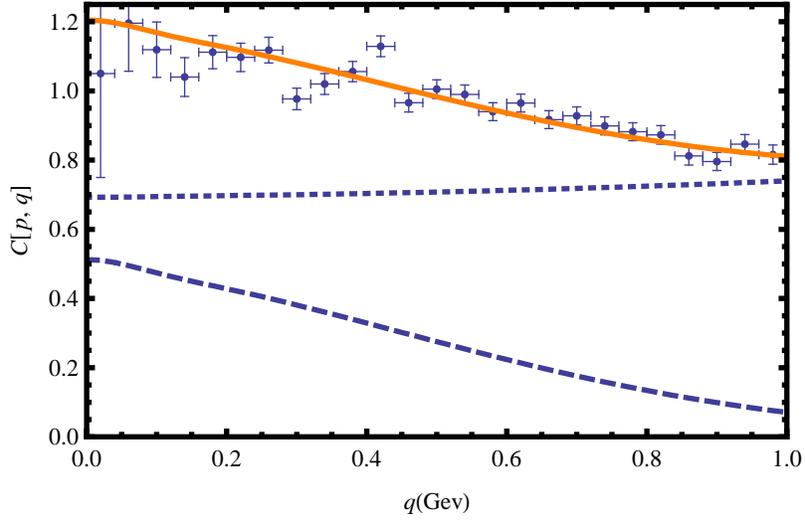}
\caption{\label{ffig5} The
 correlation functions of unlike-sign pions in the
$0.7<p_{T}<1.0$ GeV bin measured by the ALICE Collaboration  from
Refs. \cite{Alice,data} (solid dots) and those calculated from  the
analytical model (solid line). The  contributions to the
nonfemtoscopic correlation function from the first term of Eq.
(\ref{data-NF})
 (dotted line) and from the second one (dashed line) are also presented. }
\end{figure}

The results of our calculations of  the nonfemtoscopic correlation
functions $C_{NF}$   are shown in Figs. \ref{fig1}-\ref{ffig5}  in
comparison with  correlation functions  reported by the ALICE
Collaboration \cite{Alice} for different transverse momenta of pion
pairs (actually, we performed calculations for the mean value in
each bin). The auxiliary factors $\Lambda (|{\bf p}_{T}|)$ differ
from unity only slightly.\footnote{Namely, they are $0.95$, $0.96$,
$0.98$, $0.99$, $0.99$ for like-sign pions (in Figs.
\ref{fig1}-\ref{fig5}, respectively), and $0.92$, $0.92$, $0.93$,
$0.91$, $0.82$ for unlike-sign pions (in Figs.
\ref{ffig1}-\ref{ffig5}, respectively).} The data for unlike-sign
pion correlations measured by the ALICE Collaboration as well as for
the PHOJET simulations of like-sign two-pion nonfemtoscopic
correlation functions at midrapidity for the total charged
multiplicity $N_{ch} \geq 12$ bin in $p+p$ collisions at
$\sqrt{s}=900$ GeV are taken from Refs. \cite{Alice} and
\cite{data}.   Note that correlations of nonidentical pions measured
by the ALICE Collaboration, as well as the PHOJET simulations of
identical two-pion  correlation functions, demonstrate  Coulomb FSI
correlations at the lowest $q_{inv}$ bin and  peaks coming from
resonance decays. These Coulomb FSI, as well as contributions from
resonance production, are not taken into account   and so are  not
reproduced in our model.

The presented results are obtained for $M = 50$, $T_{T}=\alpha=0.65$
GeV (to minimize the number of fit parameters, we fixed
$T_{T}=\alpha$ for all calculations)\footnote{Note that with these
parameter values the mean transverse momentum $\langle p_{T}
\rangle$ is about $0.58$ GeV.}, and  the fitted values of $N$ are
different for like-sign and unlike-sign pion pairs, namely, $N^{\pm
\pm}=20$ for the former and $N^{+-}=11$ for the latter.  The
relatively high value of $M$ can be interpreted as a residual effect
on the  pion subsystem of total energy-momentum conservation in a
multiparticle production process. The relation $N^{+-} < N^{\pm
\pm}$ between fitted $N$ values means that  the magnitude of the
correlations induced by a minijet  for unlike-sign pion pairs is
higher than for like-sign ones. This  happens because in the former
there  is no local charge conservation constraint for the production
of oppositely charged pion pairs   and, therefore, one can expect
less identically charged pion pairs from the fragmenting minijets
than oppositely charged ones.

 One can see from the figures that the
behavior of the nonfemtoscopic correlation functions of pions,
$C_{NF}$, is reproduced well despite the simplicity of our model.
This is a result of the competition of the two trends: an increase
of the correlation function with $q_{inv}$ because of momentum
conservation and a decrease of it due to  fragmentation of one
minijet into the registered pion pair.  Figures \ref{fig5} and
\ref{ffig5} also demonstrate the relative contribution of the  first
and second terms in Eq. (\ref{data-NF}) to the nonfemtoscopic
correlation functions.

\section{Analytical model for event-by-event momentum spectra fluctuations}

As it follows from our previous discussion, the lower magnitude of
the nonfemtoscopic correlations at relatively low $q_{inv}$ for
like-sign pion pairs as compared to the correlations of unlike-sign
pions is natural if pions are produced through minijet
fragmentation. However,  this cannot be  the case for other
production mechanisms that do not include noticeable  production of
minijets. For example, if thermalization takes place in $p+p$
collisions and hydrodynamical evolution forms the "soft" momentum
spectra\footnote{Hydrodynamic models, perhaps, can give a reasonable
description of elementary particle collisions; see, e.g.,
\cite{Werner}. }, then  the production of minijets at relatively low
$p_{T}$ is, typically, reduced. In this case utilization of the
nonfemtoscopic correlations of like-sign pion pairs obtained in the
PHOJET and similar event generators as a correlation baseline for
the femtoscopic correlations, see Eq. (\ref{4-1}), can be in doubt.
Then the question arises whether the unlike two-pion correlations in
$p+p$ collisions, which  are reproduced in  PHOJET-like models, are
ultimately caused by minijets and conservation laws only, or whether
a similar behavior can be attributed to hydrodynamics also.

First, note that there are no correlations induced by the exact
global energy-momentum conservation in hydrodynamic models, and
corresponding conservation laws are satisfied only in average for
particles that are produced at some hypersurface where hydrodynamics
is switched off. The global energy-momentum conservation constraints
can be added on an  event-by-event basis  if the post-hydrodynamical
hadronic stage is calculated by means of some hadronic cascade model
(the so-called "hybrid" model). One more source of the
nonfemtoscopic correlations  in such models are event-by-event
fluctuations of initial conditions for the hydrodynamical stage.
These fluctuations result in fluctuations of the two-particle and
single-particle momentum spectra, and, as usual, the effect of the
fluctuations  is  more pronounced for small systems. Then, an
important question is whether such  correlations can  be similar to
minijet induced correlations that, as we know, may reproduce
unlike-sign pion correlations at relatively low $q_{inv}$.

Let us give  an illustrative example  of   nonfemtoscopic
correlations that appear because of  event-by-event fluctuating
initial conditions for the hydrodynamic stage (in hybrid models this
stage is matched  with the subsequent hadronic cascade stage) and
are similar to  the ones produced by minijets. Suppose that the
$N$-particle probability density is  defined as
\begin{eqnarray}
P_{N}(p_{1}, p_{2},...,p_{N})= \sum_{i} w(u_{i})P_{N}(p_{1},
p_{2},...,p_{N};u_{i}),
 \label{8-1}
\end{eqnarray}
where $P_{N}(p_{1}, p_{2},...,p_{N};u_{i})$ is the $N$-particle
probability density for some $u_{i}$ type of the initial conditions,
and  $w(u_{i})$ denotes the distribution over initial conditions,
$\sum_{i}w(u_{i})=1$. To analyze the possible effect of fluctuating
initial conditions,  here we neglect conservation law constraints
and the production of minijets. Because we assume uncorrelated
particle emissions for each specific initial condition, one can
write
\begin{eqnarray}
P_{N}(p_{1},
p_{2},...,p_{N};u_{i})=f(p_{1};u_{i})f(p_{2};u_{i})...f(p_{N-1};u_{i})f(p_{N};u_{i}),
 \label{20}
\end{eqnarray}
where  we normalize $f(p;u_{i})$ as
$\int\frac{d^{3}p}{E}f(p;u_{i})=1$, and then $K=1$; see  Eqs. (\ref
{11n}), (\ref {12n}). The two-particle nonfemtoscopic correlation
function $C_{NF}$ then reads
\begin{eqnarray}
C_{NF}(p_{1},p_{2})=\frac{\sum_{i}w(u_{i})f(p_{1};u_{i})f(p_{2};u_{i})}
{\sum_{i}w(u_{i})f(p_{1};u_{i})\sum_{j}w(u_{j})f(p_{2};u_{j})}.
 \label{21}
\end{eqnarray}

Evidently, the different type of   fluctuation, i.e., the form of
the  distribution $w(u_{i})$, leads to a different  behavior of the
nonfemtoscopic correlations. To illustrate that fluctuations can
lead to the nonfemtoscopic correlation functions that are similar to
the ones induced by  minijets, let us consider the toy model where
\begin{eqnarray}
w(\textbf{u}_{T})=\frac{\alpha^{2}}{\pi}
\exp(-\textbf{u}_{T}^{2}\alpha^{2}),
 \label{22-n} \\
f(p;\textbf{u}_{T})=\frac{\beta^{2}\gamma }{ \pi^{3/2}}
E\exp(-(\textbf{p}_{T}-\textbf{u}_{T})^{2}\beta^{2})\exp(-p_{L}^{2}\gamma^{2}),
 \label{23-n}
\end{eqnarray}
and normalization is chosen in such a way that $\int d^{2}u_{T}
w(\textbf{u}_{T})=1$ and $\int \frac{d^{3}p}{E}
f(p;\textbf{u}_{T})=1$. The main feature of such a model is that
event-by-event single-particle transverse momentum spectra have a
maximum for event-by-event fluctuating $\textbf{p}_T$ values. Such
momentum spectrum fluctuations could take place, e.g., in
hydrodynamics with a highly inhomogeneous initial energy density
profile without cylindrical or elliptic symmetry.\footnote{It seems
that this  is the case in heavy ion collisions, where nonsymmetrical
fluctuations of initial conditions lead  to nonzero $v_3$ and higher
flow harmonics (see, e.g., Ref. \cite{v3}).}
 One can easily see
that in this case $C_{NF}$ decreases with $q^{2}_{T}$,
\begin{eqnarray}
C_{NF}(p,q) \sim \exp (-
\frac{\beta^{4}}{2(\alpha^{2}+\beta^{2})}q^{2}_{T}),
 \label{22}
\end{eqnarray}
and  this means (after taking into account (\ref{15}) and
(\ref{15.1})) that $C_{NF}$ decreases with $q_{inv}^{2}$ too, which
is similar to the behavior of $C_{NF}$  if the nonfemtoscopic
correlations are induced by minijets. At the same time,   unlike the
latter, the hydrodynamical fluctuations lead to similar correlations
for like-sign and unlike-sign pion pairs. Then,  our analysis
suggests that, up to different resonance yields, the value of the
slope of the correlation baseline at relatively low $q_{inv}$ can be
somewhere between pure hydrodynamic (i.e., the same as for
nonidentical pion pairs) and pure minijet (i.e., lower than for
nonidentical pion pairs) scenarios.

\section{Conclusions}

We presented here a simple analytical model that takes into account
correlations induced by the total transverse momentum conservation
as well as correlations induced by the minijets. It is shown that
the model gives a reasonable description of the correlations of
nonidentical pions measured by the  ALICE Collaboration \cite{Alice}
in $p+p$ collisions at $\sqrt{s} = 900$ GeV, and also the
nonfemtoscopic correlations of identical  pions generated in the
PHOJET simulations of $p+p$ collisions and utilized by the ALICE
Collaboration \cite{Alice} as the correlation baseline. We  conclude
that the cluster (minijet) structures in the final momentum space of
produced particles can result in noticeable nonfemtoscopic two-pion
correlation functions  that decrease when $q_{inv}$ grows at
relatively low $q_{inv}$, while the global energy-momentum
conservation constraints typically result in an increase with
$q_{inv}$ for fairly high $q_{inv}$.  Our model can be utilized for
simple estimates of the nonfemtoscopic correlations induced by
minijets and conservation laws  that  contribute to the total
two-particle correlation functions.

There can be different types of multiparticle production mechanisms,
and  some of them  could result in qualitatively similar
nonfemtoscopic correlation functions. We presented  heuristic
arguments that the two-pion nonfemtoscopic correlation functions
calculated in hydrodynamics with event-by-event fluctuating initial
conditions can be qualitatively similar at relatively low $q_{inv}$
to the ones calculated in the PHOJET-like generators, where the
nonfemtoscopic correlations for low $q_{inv}$ are mainly caused by
 minijets. It is worth noting an important difference between
the nonfemtoscopic correlations induced by minijets and
hydrodynamical fluctuations: while  the former lead to a higher
magnitude of the nonfemtoscopic correlations for unlike-sign pion
pairs as compared to like-sign pions, the latter result in a similar
(up to the resonance contributions) strength of the nonfemtoscopic
correlations for identical and nonidentical pions. Then, if the
applicability of hydrodynamics to $p+p$ collisions is supported,
such an analysis allows one to estimate the correlation baseline
and, so, to extract the femtoscopic scales in these collisions by
means of tuning  the hydrokinetic model  to reproduce the
experimental unlike-sign  pion correlations.

Because the particle production mechanisms in $p+p$ are  still
unclear, the  nature of nonfemtoscopic correlations in these
collisions  is also an open question. Different dynamical models,
that reproduce unlike pion correlations, can give different
estimates of the correlation baseline and, so, lead to different
results for  the correlation femtoscopy analysis of the space-time
scales of the collision process. In our opinion,  this difficulty
can be overcome  if the correlation analysis is applied not to
femtoscopic correlation functions, but to experimental data reported
for the total correlation functions of  pion pairs. Then the
space-time scales can be estimated by means of the dynamic model
that will also be able to reproduce, among other observables, these
compete correlations.

\section{Acknowledgments}
The research  was  carried out  within the scope of the EUREA:
European Ultra Relativistic Energies Agreement (European Research
Group: "Heavy Ions at Ultrarelativistic Energies"), and is supported
by the National Academy of Sciences of Ukraine (Agreement - 2012)
and by the  State fund for fundamental researches of Ukraine
(Agreement  -  2012).

\end{document}